# IoT-Based Pothole Mapping Agent with Remote Visualization

**Umar Yahya**
Motion Analysis Research Laboratory,
Islamic University in Uganda
Kampala, Uganda
umar.yahya@ieee.org

**Mwaka Lucky**
Motion Analysis Research Laboratory,
Islamic University in Uganda
Kampala, Uganda
mwakahorus@gmail.com

**Muhammed Mansoor**
Motion Analysis Research Laboratory,
Islamic University in Uganda
Kampala, Uganda
muharaj@gmail.com

**Nankabirwa Sharifah**
Motion Analysis Research Laboratory,
Islamic University in Uganda
Kampala, Uganda
shifahmoshe7@gmail.com

**Abdal Kasule**
Department of Computer Science and
IT, Islamic University in Uganda
Kampala, Uganda
abdal78@gmail.com

**Kasagga Usama**
Department of Computer Science and
IT, Islamic University in Uganda
Kampala, Uganda
kasagga2011@gmail.com

*Abstract* – Driving through pothole infested roads is a life hazard and economically costly. The experience is even worse for motorists using the pothole filled road for the first time. Pothole filled road networks have been associated with severe traffic jam especially during peak times of the day. Besides not being fuel consumption friendly and being time wasting, traffic jams often lead to increased carbon emissions as well as noise pollution. Moreover, the risk of fatal accidents has also been strongly associated with potholes among other road network factors. Discovering potholes prior to using a particular road is therefore of significant importance. This work presents a successful demonstration of sensor-based pothole mapping agent that captures both the pothole's depth as well as its location coordinates, parameters that are then used to generate a pothole map for the agent's entire journey. The map can thus be shared with all motorists intending to use the same route.

## I. INTRODUCTION

Pothole infested roads are both a serious life hazard as well as economically costly[1][2]. Life hazards including increased risk of motorist accidents, increased carbon emissions, and traffic jam thefts have been strongly associated with the pothole infested road networks[3]. The economic cost of traffic jams induced by pothole filled roads and exit junctions is also significant as motorists end up using more fuel in addition to time wastage[2], [4], [5]. The pothole problem is further exacerbated by poor drainage systems especially during heavy downpour. Moreover, for economically constrained Countries without immediate plans to meet fix the poor roads, potholes are a strong point of frustration for prospective investors[1], [3], [6], [7].

Pothole detection research has therefore gained momentum in research years. The common pothole detection approaches utilized in previous studies include application of computer vision and image processing techniques to evaluate road surface [8]–[13], utilization of accelerometer data generated by android held devices [14]–[16], as well as vibration based systems[14]. While the results achieved by previous related research are promising, a low-cost real-time pothole detection system has not been achieved yet in literature. There is therefore a genuine need to develop an efficient crowd-sourced pothole mapping solution that will eventually benefit every motorist intending to use a pothole infested road.

With advancement in sensing technologies, it has become increasingly the norm to capture various useful data using sensors without or with minimal human involvement [17]. This work therefore presents an implementation of a wirelessly driven Robot Agent that is capable of detecting potholes during a journey. The Robot Agent not only detects the pothole, but also captures its GPS coordinates and measures its depth, both values it then continuously relays to a remote server to enable generation of a pothole map upon completion of its journey. The map can then be shared with all prospective users of the Robot's travelled route. The ultimate of aim of research presented in this current paper, is to attach the developed pothole mapping module on motor vehicles so as to facilitate continuous pothole detection automatically as they move along their respective routes. Moreover, the pothole map data will also be valuable input for various stakeholders such as road maintenance agencies as well as transport policy making bodies.

## II. METHODS

Implementation of the pothole mapping agent introduced in this current work comprises of four main modules: - Wireless Agent Control (WAC) module, pothole detection and characterization (PDC) module, pothole data relay (PDR) module, and pothole map generation (PMG) module.

Fig. 1 illustrates the overall architecture of the pothole mapping agent.

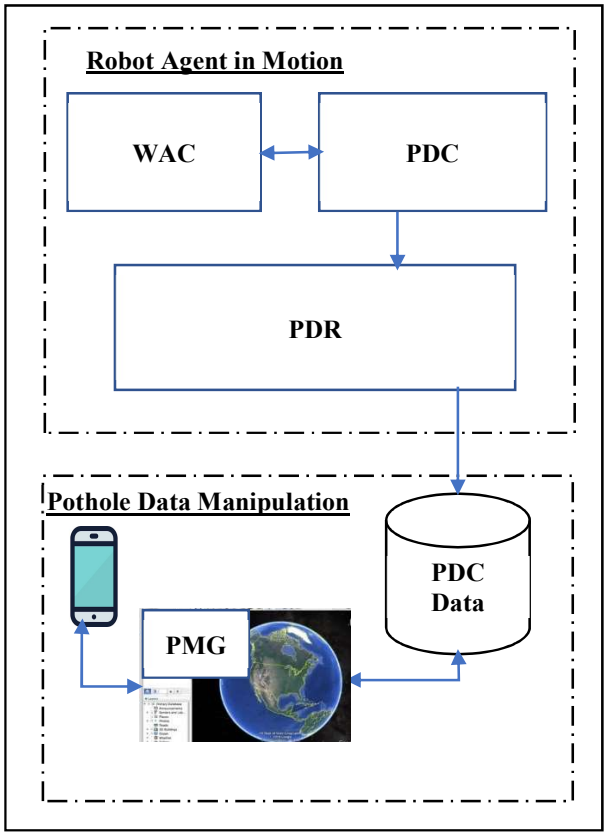

Fig. 1. Overall architecture of pothole mapping agent

### A. Wireless Agent Control (WAC) Module

The WAC module comprises of components responsible for moving the Pothole Robot Agent (PRA) pictured in Fig. 2. Components include a surface mount atmega328 Arduino Nano for managing Bluetooth connectivity and remote control, an L298M motor driver for controlling the movement and direction of the PRA, gear motor for moving the wheels of the PRA, a Bluetooth Module HC06 and facilitating the remote wireless communication during PRA movement, and an 11.1V Solar-rechargeable battery for powering all the electric components of the PRA.

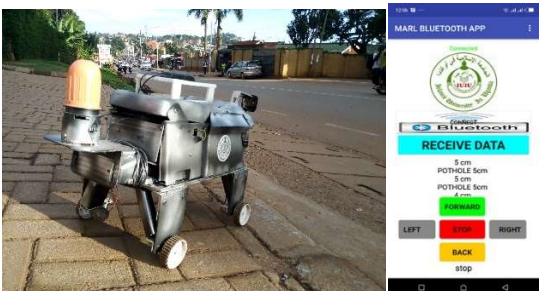

Fig. 2. Remote Controlled PRA

### B. Pothole Detection and Characterization (PDC) Module

The PDC module comprises of components responsible for detection of potholes and their characterization (i.e measuring their depth as well as capturing their geographic location coordinates) shown in Fig. 3. Pothole detection is enabled by a water proof ultrasonic sensor placed under the front-end of the PRA that continuously measures its distance from the ground and compares with its distance from the ground on a flat surface. The difference between the distances is the pothole depth. A positive distance difference implies detection of a pothole in the road, while a negative distance implies a hump has been detected in the road. The corresponding coordinates of the detected pothole or hump are simultaneously captured at the detection instance using a GPS Module Board EEPROM Antenna. Management of this interaction between the ultrasonic sensor and the GPS module is through a separate atmega328 Arduino Uno Microcontroller.

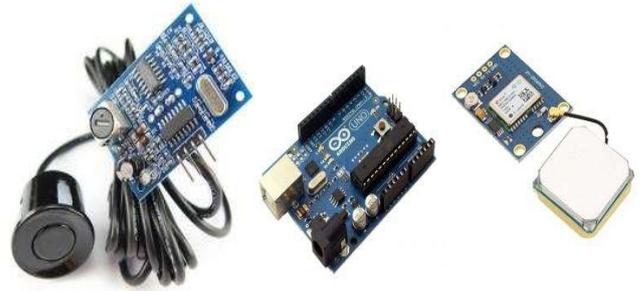

Fig. 3. PDC module components

### C. Pothole Data Relay (PDR) Module

The recorded pothole characterization data (i.e pothole location and depth) captured by the PDC module is continuously transmitted to both a remote MySQL database as well as to a registered SIM card number as an SMS, as can be seen in Fig. 4.

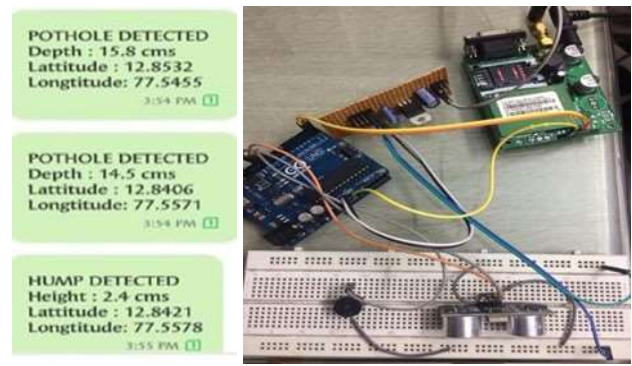

Fig. 4. SMS text with pothole characterization data sent by PDR module

### D. Pothole Map Generation (PMG) Module

The pothole characterization data received is then plotted using Google Earth to generate the pothole map for the whole journey of the PRA as seen in Fig. 5. The generated map can thus be shared among prospective road users for their convenience and better trip route planning.

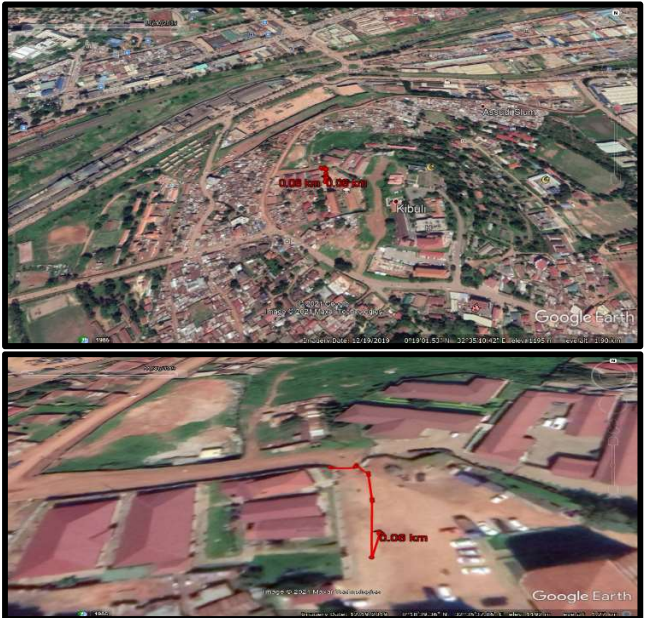

Fig. 5. Pothole map generated from GPS data received from PCD module

### III. RESULTS AND DISCUSSION

This work has presented a low-cost pothole detection and characterization system with pothole map visualization. Previous research that achieved promising pothole detection solutions require costlier computing environments compared to the solution presented in this current work [10], [12], [15], [16], [18]–[23]. Moreover, unlike in earlier works, this current research has demonstrated the feasibility of having a crowd-sourced pothole mapping system for the benefit of a wider road user community.

#### A. Automatic Real-time Pothole Detection

This work has demonstrated a feasible approach to detect and characterize potholes using low-cost sensing technologies. Extension of this current work will include a mobile application which uses the pothole characterization data to inform motorists of any road's condition whose pothole profile has ever been captured. Moreover, a pothole free road today may not be the same tomorrow, and therefore the continuous capture and sharing of pothole data for the different roads in a crowd-sourced manner will offer up-to-date pothole profiles.

#### B. Volume of Pothole Data Collected

The volume of pothole data that would be generated after a full-scale implementation of the presented system could offer further insights beyond the immediate pothole data usage as required by road users. Insights such as how long road develop potholes after repair, the most affected roads, traffic flow along the roads with persistent disturbing pothole profiles, would all be great insights in as far as policy and decision making on the state of road networks is concerned. By applying different data analytics algorithms, more meaningful insights could be mined for the huge volume of captured pothole data, similar to what has been applied during evaluation of athletes using 3D-EMG data[24]. An extension of the current work will aim to develop a comprehensive dynamic pothole information system where policy makers would be able to login in order to access insights required for making executive decisions.

### IV. CONCLUSION

This work presented a low-cost pothole mapping agent. The demonstrated success of the approach introduced in this work paves way for extended development of on-board pothole characterization modules that could be attached to motor vehicles thereby resulting in a more frequently updated data-rich crowd-source pothole mapping system. A full-scale deployment of the introduced approach could significantly reduce the huge economical and life risks associated with driving along roads without being aware of their pothole profiles. Moreover, policy makers and key decision makers on matters pertaining to road maintenance would benefit from the frequently updated status on the profile of different motorized transport networks